\documentclass[conference, letterpaper, 10pt]{IEEEtran}
\usepackage[utf8]{inputenc}
\usepackage[T1]{fontenc}

\usepackage{eulervm} 
\usepackage{dsfont} 

\usepackage[american]{babel}

\usepackage{amsmath}
\usepackage{amsthm}

\usepackage{amssymb}
\usepackage{wasysym}
\usepackage{textcomp}

\usepackage{tikz}
\usetikzlibrary{shapes, patterns, positioning}

\usepackage[hidelinks]{hyperref}

\usepackage[
	sort&compress,
	nameinlink,
	noabbrev,
        capitalise
]{cleveref}
\Crefname{algocf}{Algorithm}{Algorithms}
\crefalias{AlgoLine}{line}

\makeatletter
\let\cref@old@stepcounter\stepcounter
\def\stepcounter#1{%
  \cref@old@stepcounter{#1}%
  \cref@constructprefix{#1}{\cref@result}%
  \@ifundefined{cref@#1@alias}%
    {\def\@tempa{#1}}%
    {\def\@tempa{\csname cref@#1@alias\endcsname}}%
  \protected@edef\cref@currentlabel{%
    [\@tempa][\arabic{#1}][\cref@result]%
    \csname p@#1\endcsname\csname the#1\endcsname}}
\makeatother

\usepackage[acronym]{glossaries-extra}
\setabbreviationstyle[acronym]{long-short}

\usepackage{csquotes}

\usepackage[bibstyle=ieee, citestyle=numeric-comp, maxbibnames=5, minbibnames=5, date=year, maxcitenames=1, mincitenames=1, doi=false, isbn=false, url=false, backend=biber]{biblatex}
\DeclareFieldFormat{journaltitle}{#1\isdot}
\DeclareFieldFormat[book]{title}{{#1\isdot}}

\usepackage{multirow}
\usepackage{graphicx}
\usepackage{xcolor}
\usepackage{soul}

\usepackage{booktabs}
\usepackage{flushend}
\usepackage[
    protrusion=true,
    activate={true,nocompatibility},
    final,
    tracking=true,
    kerning=true,
    spacing=true,
    factor=1100]{microtype}
\SetTracking{encoding={*}, shape=sc}{40}

\usepackage{mathtools}

\usepackage{braket}
\usepackage{sansmath}

\usepackage{ifdraft}
\usepackage{etoolbox, xpatch}

\usepackage[normalem]{ulem}
\usepackage{tabularx}

\usepackage{makecell}
\usepackage{textcomp}
\def\nobreakbefore{%
  \relax\ifvmode\else
    \ifhmode
      \ifdim\lastskip > 0pt\relax
        \unskip\nobreakspace
      \fi
    \fi
  \fi
}
\let\oldcite\cite
\renewcommand\cite{\nobreakbefore\oldcite}
\usepackage{enumitem}

\usepackage[ruled, linesnumbered]{algorithm2e}
\SetKwInput{KwInput}{Input}
\SetKwInput{KwOutput}{Output}

\AtEveryBibitem{% Clean up the bibtex rather than editing it
 \clearname{editor}
 \clearfield{month}
  \clearfield{extra}
}
\DeclareSourcemap{
  \maps[datatype=bibtex]{
    \map{
      \step[fieldset=note, null]
    }
  }
}
\DefineBibliographyStrings{english}{%
    andothers = {et al\adddot}
}
\defbibheading{secbib}[\bibname]{%
  \section*{#1}%
  \markboth{#1}{#1}}

\usepackage{siunitx}               % A better way of writing things with units

\DeclareSIUnit\px{px}

\theoremstyle{plain}

\theoremstyle{definition}

\usepackage{caption}
\newacronym{gcd}{GCD}{Greatest Common Divisor}

\begin{document}

\title{
Optimal Qubit Reuse for Near-Term Quantum Computers\vspace{-2ex}}

 \author{

Sebastian Brandhofer,$^{1, 2}$ Ilia Polian,$^{1}$ Kevin Krsulich$^{2}$ 
\\\\
\begin{minipage}[c]{\textwidth}
\centering
\small $^1$ Institute of Computer Architecture and Computer Engineering and Center for Integrated Quantum Science and Technology, University of~Stuttgart, Stuttgart, Germany, e-mail: \{sebastian.brandhofer, ilia.polian\}@iti.uni-stuttgart.de\\
\small $^2$ IBM Quantum, IBM TJ Watson Research Center, Yorktown Heights, NY, USA, e-mail: kevin.krsulich@us.ibm.com
\vspace{-2ex}
\end{minipage}
}

\maketitle

\begin{abstract}
Near-term quantum computations are limited by high error rates, the scarcity of qubits and low qubit connectivity. Increasing support for mid-circuit measurements and qubit reset in near-term quantum computers enables qubit reuse that may yield quantum computations with fewer qubits and lower errors.

In this work, we introduce a formal model for qubit reuse optimization that delivers provably optimal solutions with respect to quantum circuit depth, number of qubits, or number of swap gates for the first time. This is in contrast to related work where qubit reuse is used heuristically or optimally but without consideration of the mapping effort. We further investigate reset errors on near-term quantum computers by performing reset error characterization experiments. Using the hereby obtained reset error characterization and calibration data of a near-term quantum computer, we then determine a qubit assignment that is optimal with respect to a given cost function. We define this cost function to include gate errors and decoherence as well as the individual reset error of each qubit. 

We found the reset fidelity to be state-dependent and to range, depending on the reset qubit, from 67.5\% to 100\%  in a near-term quantum computer.
We demonstrate the applicability of the developed method to a number of quantum circuits and show improvements in the number of qubits and swap gate insertions, estimated success probability, and Hellinger fidelity of the investigated quantum circuits.
\end{abstract}

\maketitle

\section{Introduction\vspace{-1ex}}
Quantum computing promises significant speedup for problems in cryptography \cite{43} and chemistry \cite{9}.
However, in near-term quantum computers, the greatest challenges of effective quantum computations are a lack of qubits and the corruption of quantum states due to errors.
The limited connectivity of qubits in contemporary quantum computers aggravates these challenges further \cite{12, 8, 31, 41, 20} as additional operations, e.g. swap gates \cite{42, 12, 20, 40, 38, 13}, need to be inserted.
While the technology enabling quantum computing continuously progresses 
\cite{14}, optimizing the executed quantum computations, i.e. quantum circuits, improves the performance and can extend the computational reach of near-term quantum computers \cite{44, 1, 29, 36}.

The reset operation combined with mid-circuit measurements offers a venue for quantum circuit optimization by allowing to reuse a qubit of a quantum computer after the computations on a previously assigned quantum circuit qubit have concluded. 
The reset operation and mid-circuit measurements are essential for quantum error correction protocols \cite{24, 39} and have recently started to be supported with increased fidelity by quantum computer vendors such as IBM, Google and Honeywell \cite{25, 15, 6}.
In general, reusing qubits offers the following improvements to a given input quantum algorithm:
\begin{itemize}
    \item Reduction of the qubit count requirement of a quantum algorithm; A quantum algorithm that requires $n$ qubits may be transformed into a quantum algorithm that requires $n-i$ qubits at the cost of a longer computation duration.
    \item 
    Simplified connectivity requirements of a quantum algorithm; A reused qubit may require less effort than a new qubit to satisfy the required connectivity.
    \item Reduction of error; First, less stringent connectivity requirements lead to fewer errors \cite{5, 13, 35, 34}. Furthermore, the quality of qubits in near-term quantum computers varies significantly \cite{3, 28, 4}---a subset of qubits incurs less error. Reducing the required amount of qubits may allow to avoid qubits with a higher error rate.
\end{itemize}
However, reusing qubits may come at the cost of increased quantum circuit duration, hence potentially increasing the errors due to decoherence.
Paired with high variability in reset operation error (see this work in \cref{sec char_res}), this requires to carefully optimize the quantum circuit with qubit reuse for improving the result quality yielded by near-term quantum computers.
The quantum circuit optimization method developed in this work improved the Hellinger fidelity by up to 4.3x on the near-term quantum computer \texttt{ibm\_hanoi} and replaced swap gate insertions by qubit reuse at minimal quantum circuit depth.
The paper at hand addresses the optimization of quantum circuits via qubit reuse by:
\begin{itemize}
    \item Characterizing and reporting the error of reset operations on a 27-qubit quantum computer for the first time while considering state-dependent errors and errors due to concurrent reset operations. 
    \item Introducing a novel SAT-based quantum circuit optimization method that determines the optimal application of swap gate insertions and qubit reuse to produce a quantum circuit conforming to a target quantum computer.
    \item Combining the heavy-weight but optimal SAT-based approach with an efficient at-runtime qubit assignment \cite{4, 28} based on the previous reset error characterization.
    \item Optimizing the number of reset repetitions for each qubit depending on the obtained reset error characterization.
    \item Evaluating optimized quantum circuits on IBM quantum computers to investigate the improvement in fidelity and quantum circuit characteristics yielded by the developed quantum circuit optimization method.
\end{itemize}
\begin{figure*}[ht!]
    \centering
    \includegraphics[width=1.0\textwidth]{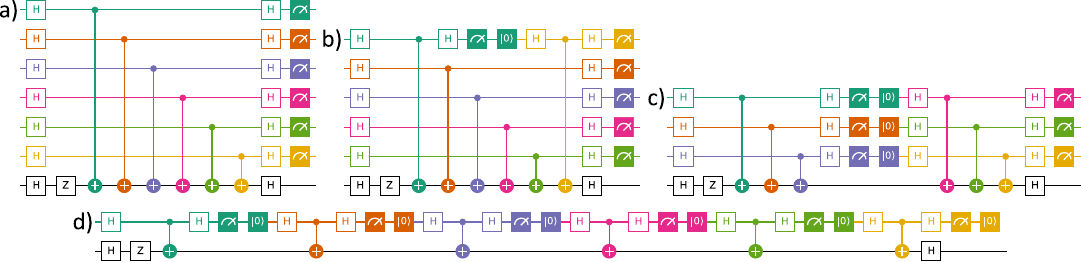}

    \caption{Quantum circuit optimization of a) a 7-qubit Bernstein-Vazirani quantum circuit with qubit reuse where, b) one qubit was reused, c) swap gates are not necessary due to qubit reuse, and d) the maximum of qubits (all but two) are reused.}
    \label{fig qubit reset reuse}
\end{figure*}

The remainder of this work is organized as follows. \cref{sec background} introduces the background for this work and \cref{sec related} outlines the related work. \cref{sec char} describes the experiments used for reset error characterization and evaluates them on an IBM quantum computer. Then, the novel quantum circuit optimization method and qubit assignment are introduced in \cref{sec method}. The introduced quantum circuit optimization method is evaluated in \cref{sec results} and the work is concluded in \cref{sec conclusion}.

\section{Qubits, Reset and Reuse} \label{sec background}

Depending on an external signal, an $n$-qubit quantum computer can manipulate, store and measure an $n$-qubit state given by:
\begin{equation}
    \ket{\psi} = \sum_{x\in\{0, 1\}^{n}} \alpha_{x} \ket{x},
\end{equation}
where $\alpha_x$ are complex probability amplitudes, i.e. they represent probabilities with $\sum_{x} |\alpha_{x}|^2 = 1$.
Measuring the complete state $\ket{\psi}$ in the standard basis yields a measurement outcome $k$ corresponding to the basis state $\ket{k}$ with probability $|\alpha_{k}|^{2}$; the state $\ket{\psi}$ collapses to the basis state $\ket{k}$.
When a part of the quantum state is measured, the quantum state partially collapses according to the yielded measurement outcome.
Before reusing a qubit, the qubit must be in a known and expected state---typically the $\ket{0}$ state.
Yielding the $\ket{0}$ state on a previously used qubit in an unknown state is achieved using the reset operation.

As noted in \cite{22}, the state of a qubit is only required between the initialization and measurement of the qubit.
This gives rise to the optimization of quantum circuits as shown in \cref{fig qubit reset reuse} where the qubit state is reset after interactions on that qubit are concluded and a measurement is performed.
The reset qubit is then able to store the state and perform the interactions of a different qubit in the quantum circuit.
These optimizations require fast high-fidelity reset and mid-measurement operations. 

In \cref{fig qubit reset reuse}, a 7-qubit Bernstein-Vazirani (BV) quantum circuit is optimized using qubit reuse.
As visible in \cref{fig qubit reset reuse}, there are multiple options for using qubit reset that improve the target quantum circuit in different ways.
Not using the qubit reuse optimization, as displayed in \cref{fig qubit reset reuse}a), yields the quantum circuit with the smallest depth if all-to-all connectivity is available.
However, with the limited qubit connectivity of near-term quantum computers, swap gates need to be inserted into the quantum circuit, further increasing the quantum circuit depth.
For instance, three swap gates would be required for quantum circuit \cref{fig qubit reset reuse}a) on the heavy-hex qubit connectivity of IBM quantum computers.

In \cref{fig qubit reset reuse}b), one qubit is reused by applying one reset and mid-circuit measurement operation. This reduces the number of swap gates in the quantum circuit by one (assuming heavy-hex qubit connectivity) while not increasing the depth of the quantum circuit.

The optimization option depicted in \cref{fig qubit reset reuse}c) reuses three qubits and does not require swap gates assuming heavy-hex qubit connectivity.
However, the quantum circuit depth is increased by 40\% compared to \cref{fig qubit reset reuse}a).

The qubit reuse option depicted in \cref{fig qubit reset reuse}d), requires the least amount of qubits and no swap gates even at linear connectivity. However, the quantum circuit depth is also significantly increased.

This example demonstrates the degrees of freedom when improving quantum circuits using qubit reuse and sets the stage for the remainder of the work.
A quantum circuit optimization method based on qubit reuse must allow optimizing for quantum circuit depth, inserted swap gates and number of qubits.

\section{Related Work} \label{sec related}
A number of works investigate the optimization of quantum circuits with regards to the minimization of swap insertions using heuristic \cite{38, 40, 35} and optimal methods \cite{42, 13}.
Furthermore, the variety of hardware modalities used in the nascent field of quantum computing gives rise to further optimizations using operations native to a specific hardware modality \cite{29, 30}.

As quantum computers support mid-circuit measurements and qubit resets with increasing fidelity \cite{25, 24, 0_, 6, 15}, recent approaches investigate quantum circuit optimization through exploiting qubit reuse \cite{27, 7, 11, 22}.
These works can be divided into highly-scalable heuristics that are able to target large-scale quantum circuits \cite{27, 7, 22} and optimal less-scalable SAT-based approaches to qubit reuse \cite{11} where the focus is in general on investigating optimization opportunities given by qubit reset and on providing a baseline for scalable heuristic approaches.

In this work, we developed an optimal approach to qubit reuse that, unlike in recent optimal approaches \cite{11}, simultaneously considers the quantum circuit depth and the mapping effort, quantified by the number of required swap gates.
Thus, the method developed in this work produces a quantum circuit that considers the connectivity of a quantum computer and can therefore directly be executed on the quantum computer.
In \cite{11}, a swap gate insertion algorithm \cite{38, 40, 42, 12, 13} must be performed independently, thus inserting swap gates that could have been addressed by qubit reuse.
In contrast to heuristic approaches \cite{27, 7, 22}, this work optimally improves a quantum circuit using qubit reuse with respect to quantum circuit depth, the number of inserted swap gates, or the number of required qubits.

The characterization of errors in quantum computers has been investigated in numerous works \cite{18, 16, 23, 17, 19, 21, 32, 33, 10, 25, 24}.
The reset error of single qubits in IBM Quantum computers has been investigated in \cite{25, 24}.
In \cite{10}, the reset operation on 5-qubit quantum computers has been characterized from a security point of view where the focus lies on secure reset operations that do not leak information to an attacker.
In this work, we characterize and report reset errors on a 27-qubit quantum computer for the first time while investigating state-dependent errors and errors due to concurrent reset operations.

\section{Reset Characterization Experiments} \label{sec char}
\label{sec char_res}
\begin{figure}
    \centering
    \includegraphics[width=1\linewidth]{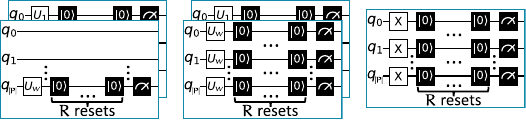}
    \caption{The three reset error characterization experiments investigated in this work, ordered by the number of required quantum circuits.
   }
    \label{fig reset_error_characterization}
\end{figure}

In this work, we perform three types of characterization experiments to quantify the per-qubit variability, the impact of the initialized state of the qubit, and the number of reset repetitions on the reset error.
\cref{fig reset_error_characterization} gives an overview of the three 
experiments.
For the first two reset error characterization experiments, $W$ random single-qubit gates $U_{1}, ..., U_{W}$ are drawn for each repetition of the characterization experiments.
In the first 
experiment, each random single-qubit gate, $r \in \{1,..., R\}$ reset operations, and a measurement operator are applied successively to each qubit $q\in P$ of the quantum computer.
In the second experiment, the same operations are applied simultaneously to each qubit of the quantum computer, rather than sequentially.
In the third characterization experiment, the random gates are replaced by the single-qubit Pauli-$X$ quantum gate and, as in the second experiment, applied simultaneously on all qubits of the quantum computer.
For each of these characterization experiments, the reset error is derived from the frequency of non-zero measurement outcomes.
The first reset error characterization experiment (left in \cref{fig reset_error_characterization}) requires $|P| \cdot W \cdot R$, the second (middle in \cref{fig reset_error_characterization}) requires $W\cdot R$ and the third (right in \cref{fig reset_error_characterization}) requires $R$ quantum circuit executions, where $|P|$ is the number of qubits on the device, $R$ is the maximum considered number of successively applied reset operations for one reset and $W$ is the maximum number of evaluated random gates.
 
We expect state-dependent reset errors to be observable in this set of characterization experiments.
First, through the random state preparations performed by the random single-qubit gates and second through any difference in outcome between the first two and the third characterization experiments.
Furthermore, we investigate the impact of errors due to concurrent reset operations by comparing the first characterization experiment, where operations and measurements are applied individually on each qubit, to the second characterization experiment where all operations are applied simultaneously on all qubits of the quantum computer.
While the third reset characterization experiment is much more efficient, i.e. requires fewer quantum circuit executions than the first two experiments, we will validate whether this experiment is sufficient to quantify the individual reset error of a qubit. 

In the remainder of this section, the introduced reset error characterization experiments are conducted on the 27-qubit quantum computer \texttt{ibmq\_ehningen}.
To counteract a bias due to shifting error rates during reset error characterization, the reset error characterization experiments are interspersed, i.e. the first reset error characterization experiment is performed with $r\in R$ reset repetition and random single-qubit gate $U_{1},..., U_{W}$, followed by the second and third reset error characterization experiment using the same experimental parameters before continuing with the next set of parameters.
We investigated the impact of up to $R=5$ reset repetitions on the reset fidelity and performed $W=50$ arbitrary state initializations.

\subsection{Comparison of Reset Characterization Experiments} \label{sec char res exp}

\begin{figure*}
    \centering
    \includegraphics[width=\linewidth]{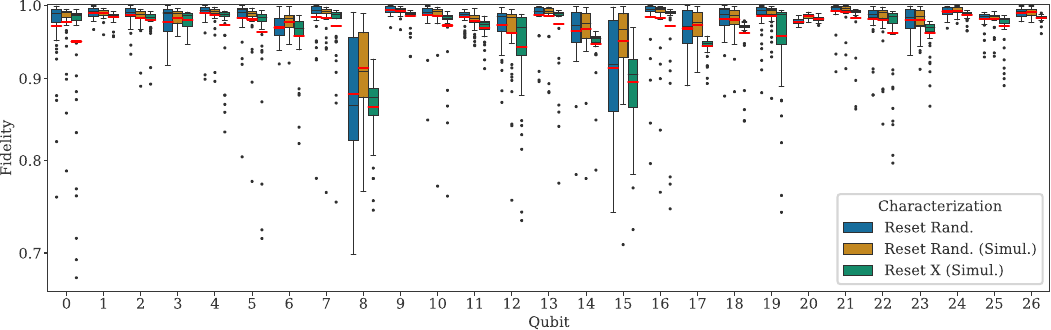}
    
    \caption{Fidelity of individually resetting an arbitrary single-qubit state (\texttt{Reset Rand.}), simultaneously resetting all qubits after preparing each of them in an arbitrary single-qubit state (\texttt{Reset Rand. (Simul.)}) and simultaneously resetting all qubits after preparing each of them in the $\ket{1}$-state  (\texttt{Reset X (Simul.)}). A red line marks the respective mean value. Experiments were conducted on the 27-qubit quantum computer \texttt{ibmq\_ehningen}.}
    \label{fig reset exps}
\end{figure*}

In \cref{fig reset exps}, the fidelity of resetting one qubit of the 27-qubit quantum computer \texttt{ibmq\_ehningen} by applying one reset operation is depicted on the y-axis while the x-axis indicates the respective qubit.
A red line marks the average reset fidelity for a qubit and reset error characterization experiment.
Except for qubits 8, 12, 15, 19, and 22, the reset fidelity shows little variance apart from the respective outliers (depicted as gray circles).
Qubit 24 demonstrated the best average reset fidelity of 98.98\% while qubit 20 shows the best worst-case behavior with a minimal reset fidelity of 96.6\% for all reset characterization experiments.
Qubit 21 was the only qubit that reached a reset fidelity of 100\% occasionally during our experiments.
Qubit 8 exhibited the worst average reset fidelity of 88.57\% while the worst overall reset fidelity of 67.55\% was reported for qubit 0, which also demonstrated the largest rate of outliers (18\%).
More than half of the qubits showed an outlier rate of 8.6\% or higher.

In general, individual qubit preparation of arbitrary state, reset, and measurement (\texttt{Reset Rand.}) yields the highest best-case reset fidelity followed by simultaneous resets of arbitrary states (\texttt{Reset Rand. (Simul.)}) and simultaneous resets of the $\ket{1}$-state.
For most qubits, the reset characterization \texttt{Reset Rand.} performs better or as well as for \texttt{Reset Rand. (Simul.)} with $\ket{1}$-state reset characterization \texttt{Reset X (Simul.)} performing worst.
However, the average reset fidelity is slightly better for the characterization experiment \texttt{Reset Rand. (Simul.)} at 97.7\% (+-3.2\%) compared to the characterization experiment \texttt{Reset Rand.} with 97.5\% (+-3.8\%).

We concluded that reset error due to concurrent reset operations is negligible for quantum circuit optimization and that the $\ket{1}$-state reset characterization \texttt{Reset X (Simul.)} can serve as a quick reset fidelity lower bound for quantum circuit optimization purposes.

\subsection{Impact of Reset Repetitions on Fidelity}
\begin{figure*}
    \centering
    
    \includegraphics[width=\linewidth]{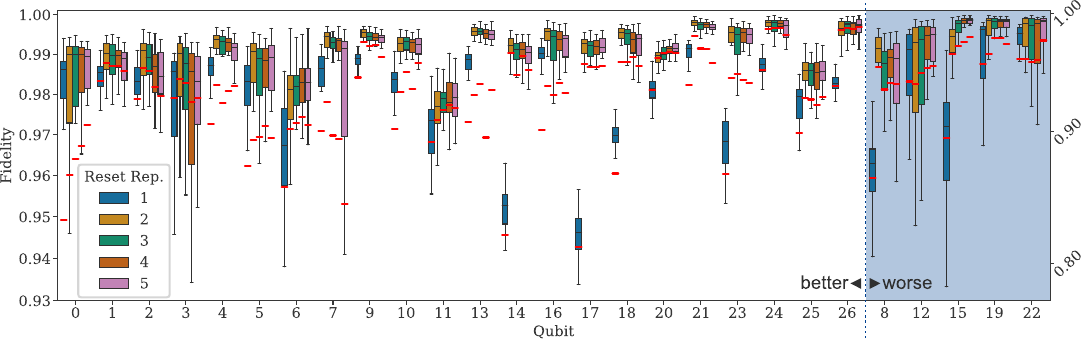}

    \caption{Fidelity of simultaneously applying the reset operation on the qubits of the 27-qubit quantum computer \texttt{ibmq\_ehningen} one to five times, after preparing each qubit in the $\ket{1}$-state (\texttt{Reset X (Simul.)}, outliers omitted).
    The reset fidelity on qubits 8, 12, 15, 19, 22 is plotted on the second y-axis. A red line marks the respective mean value.}
    \label{fig reset reps}

\end{figure*}

\cref{fig reset reps} shows the impact of repeated reset operations on the reset fidelity (y-axis) reported by the \texttt{Reset X (Simul.)} characterization experiment for each qubit (x-axis) of the \texttt{ibmq\_ehningen} quantum computer.
The reset operation is applied one to five times on each qubit, outliers are omitted.
The reset fidelity of qubits 8, 12, 15, 19, and 22 are registered on the second y-axis and ranges from 78\% to 99.9\% while the remaining qubits had a reset fidelity ranging from 93.3\% to 100\%.
For more than 80\% of the qubits, the worst reset fidelity can be observed at one single reset operation application.
Repeating the reset operation once improves the fidelity dramatically, even halving the reset infidelity for some qubits such as qubit 17.
Increasing the reset repetition further does not improve the reset fidelity for all qubits.
For instance, qubit 23 exhibits the largest reset fidelity at 3 reset operation applications with a diminishing at further reset repetitions.

We conclude that the reset fidelity varies significantly per qubit with each qubit having an optimal number of reset repetitions on average.
Thus, it appears essential for the application of qubit reuse in near-term quantum computers to employ a low-latency adaptation of a mapped quantum circuit to up-to-date reset fidelities and to select the optimal number of reset repetitions for each qubit.
\subsection{Impact of Initialized State on Reset Fidelity}
\begin{figure*}
    \centering
    \includegraphics[width=\linewidth]{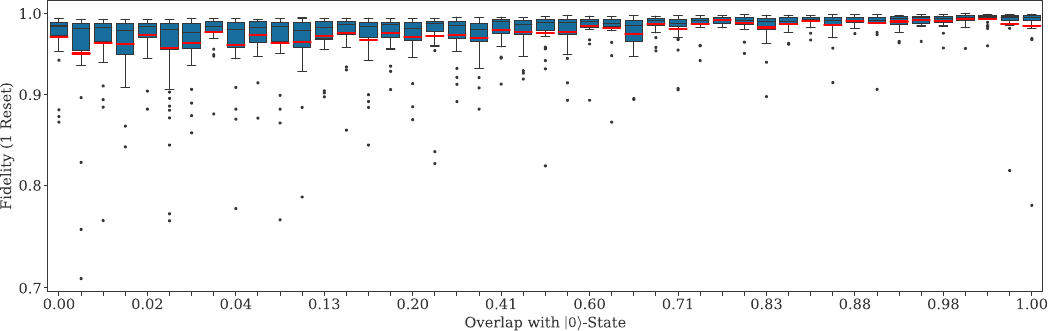}
    
    \caption{The fidelity of resetting a single-qubit state by applying one reset operation for various overlaps with the $\ket{0}$-state on the quantum computer \texttt{ibmq\_ehningen} (\texttt{Reset Rand. (Simul.)}). A red line marks the respective mean value.}
    
    \label{fig reset state}
\end{figure*}

In \cref{fig reset state}, we investigate the impact of the qubit state on reset fidelity.
Here, the (\texttt{Reset Rand. (Simul.)}) characterization experiment was conducted on all qubits of the \texttt{ibmq\_ehningen} quantum computer.
The y-axis shows the reset fidelity of all qubits as a boxplot for an initialization state with an $\ket{0}-$state overlap that is indicated on the x-axis.

The reset fidelity ranges from roughly 70\% for states that have almost no overlap with the $\ket{0}$-state to 100\% for initialization states that strongly overlap with the $\ket{0}$-state.
For an increasing $\ket{0}$-state overlap, the reset fidelity of all qubits increases gradually while the variance of the reset fidelity between different qubits decreases significantly.
Furthermore, the number and magnitude of reset fidelity outliers decrease.
The correlation between the overlap with the $\ket{0}$-state and the reset fidelity is largest when only one reset operation is applied.
This is quantified by a Pearson correlation coefficient of 0.33 that halves to 0.15 for two reset applications and further reduces to 0.13 for the maximum of five considered reset applications.

We conclude that state-dependent reset errors occur on near-term quantum computers.
These state-dependent errors can be reduced by repeatedly applying the reset operation.

\section{Quantum Circuit Optimization through Qubit Reuse} \label{sec method}
\begin{figure}
    \centering
    \includegraphics[width=0.8\linewidth]{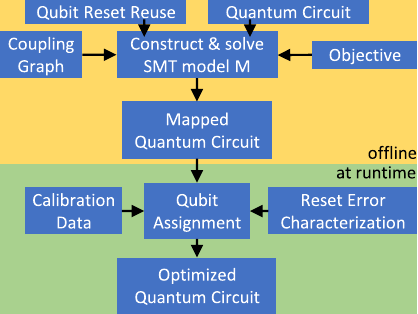}
    
    \caption{The individual steps of the developed SAT-based quantum circuit optimization method exploiting qubit reset-reuse and mid-circuit measurements.
   }
    \label{fig method_flow}
\end{figure}

\cref{fig method_flow} describes the individual steps of the developed quantum circuit optimization method that is utilizing reset operations to reuse a qubit at runtime.
The method is divided into an offline part (colored orange) which can be performed well ahead of the intended quantum circuit execution time and an at-runtime part (colored green) which is performed shortly, within minutes or seconds, before the execution of the quantum circuit.
This division into an offline and at runtime part follows the approach introduced in \cite{4}.

During the offline part, the quantum circuit can be optimized heavily, e.g. with complete methods, at a high runtime cost without necessarily incurring a delay during the at-runtime phase.
Here, a satisfiability modulo theories (SMT) model $\mathcal{M}$ is derived from the input quantum circuit, an optimization objective function, a swap insertion model (e.g. \cite{42}), and the optimizations available through qubit reset operations and subsequent qubit reuse.
The SMT model $\mathcal{M}$ is then input to the Z3-SMT solver \cite{45} that computes a mapped quantum circuit, i.e. a quantum circuit that satisfies the connectivity requirements given by the coupling map of the quantum computer, is derived from a satisfying assignment to the model $\mathcal{M}$.
These optimizations are conducted with respect to static quantum circuit properties that are not changed by a re-calibration of the quantum computer.
The three properties investigated in this work are quantum circuit depth, and the number of required swap gates or qubits.

During the at-runtime part, the quantum circuit is quickly adapted to the current error characteristics of the target quantum computer before it is computed.
The error characteristics are given by error characterization experiments (see \cref{sec char}) and calibration data \cite{2} obtained by the operator of the quantum computer e.g. IBM Quantum.
In this work, the quick adaptation is performed by using a custom cost function in conjunction with the method introduced in \cite{28}, where a mapped quantum circuit is efficiently placed onto a subset of qubits on the quantum computer depending on a given cost function.

In the remainder of this section, we will describe the construction of the SMT model (see \cref{sec sat}) and the performed qubit placement (see \cref{sec mapomatic}) in detail.

\subsection{SAT-Based Quantum Circuit Mapping with Qubit Reuse} \label{sec sat}
The constructed SMT model $\mathcal{M}$ is derived from the input quantum circuit, the ability to exploit reset operations for quantum circuit optimization, an optimization objective function and the swap insertion model $\mathcal{S}$ introduced in \cite{42}.
In fact, the model $\mathcal{M}$ is a union of the swap insertion model $\mathcal{S}$ and the qubit reset-reuse model $\mathcal{R}$ introduced in this section.
The swap insertion model $\mathcal{S}$ represents the resolution of connectivity requirements imposed by the coupling graph of the quantum computer by using swap gates.
The qubit reset-reuse model $\mathcal{R}$ represents qubit reuse opportunities after the computation on a qubit has been completed.

Essentially, the model $\mathcal{R}$ must extend typical swap insertion models \cite{42, 13} with the notion of 'unassigned' qubits.
Swap insertion models represent changes to the qubit assignment by the insertion of one or more swap gates.
For this, each qubit in the quantum circuit has an initial assignment to a qubit of the quantum computer.
This assignment changes through swap gates whenever a two-qubit gate occurs that requires interactions not included in the coupling graph of the quantum computer.
Model $\mathcal{R}$ extends this notion of qubit assignment~by:
\begin{itemize}
    \item Allowing a qubit in the quantum circuit to start without an assignment to a qubit on the quantum computer.
    \item Not applying swap gates on unassigned qubits.
    \item Enforcing the change of the assignment status of a qubit through the reset operation: The qubit on which a reset operation was applied becomes unassigned while a previously unassigned qubit becomes assigned.
    \item Only allowing reset operations after a qubit in the quantum circuit has been measured.
\end{itemize}
This requires a set of new model variables and model constraints explained in the remainder of this section.

Constructing and solving model $\mathcal{M}$ that combines a swap insertion model with the qubit reset-reuse model allows to optimally determine the number of swap gate insertions required for a target quantum circuit if qubit reset-reuse is available.
The combined model also allows to optimally trade-off the usage of swap gates and reset operations for different costs of reset operations and swap gates.
In this work, the reordering of quantum gates is not considered \cite{99}.

\subsubsection{Model Variables}
The developed qubit reset-reuse model $\mathcal{R}$ for a quantum circuit with a set of qubits $Q$ in the quantum circuit, a set of quantum gates $G$, a set of physical qubits on the quantum computer $P$, a set of measurement operators $M\subseteq G$ and the maximum considered depth $T$ contains variables:
\begin{itemize}
    \item $A = \{a_{1, 1}, ...,a_{|Q|,T}\}:$ the set of assignment statuses of a qubit, i.e. $a_{q, t}$ evaluates to true if the qubit $q$ in the quantum circuit is assigned to a physical qubit of the quantum computer at time step $t$ else the variable evaluates to false.
    \item $L = \{l_{1}, ..., l_{|M|}\}:$ the set of potential reset locations i.e. $l_{m}$ evaluates to true if a reset operation is inserted after the operation $m\in M$ else the variable evaluates to false. In this work, reset operations are only inserted after a measurement operator.
\end{itemize}
The maximum depth $T$ is determined by first applying a heuristic swap insertion algorithm \cite{38, 40, 35}.
Then, inspecting the worst-case quantum circuit depth if the maximum number of qubit reuses were to be applied during the optimization of the input quantum circuit.
As the complete and optimal swap insertion model $\mathcal{S}$ would always yield a better solution than the heuristic swap insertion algorithm, we can use the determined depth as an upper bound $T$.

The following variables in the swap insertion model $\mathcal{S}$ are interacting with variables in the qubit reset-reuse model $\mathcal{R}$, i.e. they occur in the same constraints:
\begin{itemize}
    \item $\Pi = \{\pi_{1, 1, 1},...\pi_{|Q|, |P|, T}\}:$ is the set of qubit assignments, i.e. iff the variable $\pi_{q, p, t}$ evaluates to true in a time step $t\in \{1, ..., T\}$, a qubit in the quantum circuit $q\in Q$ is mapped to the physical qubit $p\in P$.
    \item $E = \{\sigma_{1, 1, 1}, ..., \sigma_{|P|, |P|, T}\}:$ is the set of swap insertions, i.e. iff the variable $\sigma_{i, j, t}$ evaluates to true, a swap gate is inserted between physical qubits $i\in P$ and $j \in P$ at time step $t\in \{1, ..., T\}$. Consequently, the corresponding qubit assignment variables $\pi_{\_, i, t}$ and $\pi_{\_, j, t}$ must also change.
    \item $X_{g}$ is the set of gate locations for quantum gate $g$, e.g. single qubit gates need to be executed on a physical qubit. Thus, for single qubit gates $X_{g} = \{x_{g, 1}, ..., x_{|G|, |P|}\}$ with variable $x_{g, t}$ evaluating to true iff gate $g\in G$ is executed at time $t\in \{1, ..., T\}$.
    \item $D_{g} = \{d_{1,1}, ..., d_{|G|, T}\}$ is the set of gate timings, i.e. variable $d_{g, t}$ evaluates to true iff gate $g\in G$ occurs at time step $t\in \{1, ..., T\}$.
\end{itemize}

\subsubsection{Model Constraints}
Initially, the assignment status variables $a_{\cdot, 1}$ are not constrained for the first time step and can thus be set arbitrarily.
For subsequent assignment status variable changes at time step $t\in \{2, ..., T\}$, we require a corresponding reset operation to occur.
\begin{equation}
    \left( \neg a_{q, t} \wedge a_{q, t+1} \right) \rightarrow \bigvee_{m\in M, p\in P}\!\!\!\!\!\!\!\left( l_{m} \wedge x_{m, p}\wedge d_{m,t} \wedge \pi_{q, p, t+1}\right),
\end{equation}
where $\neg, \vee, \wedge, \rightarrow$ are the logical negation, disjunction, conjunction and implication operations.
This constraint enforces that whenever a previously unassigned qubit $q\in Q$ becomes assigned at time step $t+1$, a reset must happen at time step $t$ and on a physical qubit $p\in P$ to which the qubit $q$ is assigned at time step $t+1$.

Reciprocally, if a qubit $q\in Q$ becomes unassigned at time step $t+1$, constraints enforce that a corresponding reset happens at time step $t$ (and vice versa): 
\begin{equation}
    \left( a_{q, t} \wedge \neg a_{q, t+1} \right) \leftrightarrow \left(l_{m_{q}} \wedge d_{m_{q}, t}\right)
\end{equation}
where $m_{q}\in M$ is the measurement operator that acts on a qubit $q\in Q$.

An unassigned qubit $q\in Q$ cannot participate in the computation of the quantum circuit, which yields a set of consequences.
First, quantum gates can not act on unassigned qubits as they have not been mapped to a physical qubit on the quantum computer:
\begin{equation}
    \neg a_{q, t} \rightarrow \neg d_{g, t},
\end{equation}
 where $g\in G$ is a quantum gate that acts on the quantum circuit qubit $q\in Q$.

The swap insertion model $\mathcal{S}$ must further be modified to allow for changes to qubit assignment variables $\pi_{q, p, t}$ using reset operations instead of only using swap gates \cite{42}.
Furthermore, the qubit assignment variables $\pi_{q, p, t}$ are not necessarily injective anymore as two quantum circuit qubits $q, q' \in Q$ may be mapped to the same physical qubit $p\in P$ as long as one of the qubits $q, q'$ are unassigned.
The used swap insertion model must be adapted to permit these non-injective qubit assignments.

\subsubsection{Evaluated Objective Functions}
In this work, we evaluated three objective functions that guide satisfiable assignments to variables in model $\mathcal{M}$ towards preferred quantum circuits that we expect to perform better on a quantum computer.

The first optimization objective investigated minimizes the total depth of the quantum circuit:
\begin{equation} \label{eq obj depth}
    \min Z, Z\geq t\wedge d_{g, t}, \forall g\in G, \forall t\in\{1, ..., T\},
\end{equation}
where $Z$ is an integer variable.
The quantum circuit depth is correlated with the duration of a quantum circuit and hence with incurred decoherence errors.
Minimizing the quantum circuit depth, therefore, is expected to yield a quantum computation result with fewer errors.

Another metric to be minimized using qubit reset-reuse is the number of qubits on the quantum computer required to execute a quantum circuit.
We, therefore, formulate the following objective function
\begin{equation} \label{eq obj qubits}
   \min \sum_{p\in P} \left( \vee_{q\in Q, t\in \{1, .., T\}} \pi_{q, p, t} \right),
\end{equation}
where the number of qubits required by a target quantum circuit is determined by the qubit assignment variables $\pi_{q, p, t}$, i.e. each physical qubit $p$ that a qubit in the quantum circuit $q$ is assigned to at least once counts towards the number of required qubits on the quantum computer.

Finally, the number of inserted swap gates can be reduced using qubit reset-reuse by minimizing
\begin{equation}\label{eq obj swap}
    \sum_{e\in E} e,
\end{equation}
where $E$ is the set of swap insertions.
\subsubsection{Deriving the Optimized Quantum Circuit}
A satisfiable assignment to the variables in model $\mathcal{M}$ yields a valid quantum circuit, i.e. each physical qubit on the quantum computer is used at most once for the computation of a quantum gate during a specific time step.
Furthermore, any connectivity requirements of the quantum circuit that are not directly satisfied by the coupling graph of the quantum computer are resolved by either inserting swap gates or inserting reset operations into the quantum circuit.
The inserted quantum gates can be derived by inspecting the changes to qubit assignment variables $\pi_{q, p, t}$, qubit assignment status variables $a_{q, t}$ and inserted reset operations $l_{m}$.
\RestyleAlgo{ruled}
\begin{algorithm}
	\caption{Qubit assignment with reset operations}
         \label{alg:qubit_assignment}

	\KwInput{A mapped quantum circuit $C$, the coupling graph of the target quantum computer $H$}
	\KwOutput{A mapped quantum circuit with optimized qubit assignment}
	\Begin{

        $\xi_{g_x} \gets$ retrieve gate fidelities from calibration\;        
        
        $\xi_{r, p} \gets$ determine adapted reset fidelity from characterization\;
        
        $R_{p} \gets$ determine the number of repetitions for one reset operation per qubit from $\xi_{r, p}$\;
        $\bar{A} \gets \texttt{subgraph\_isomorphism}$(graph($C$), $H$)\;
		\For{$\bar{a} \in \bar{A}$}{
            $K_{\bar{a}} \gets  \text{cost}(\bar{a}, \xi, R_{p})$\;
        }
		$\bar{a}_{f} \gets \bar{a}$ with lowest cost $K_{\bar{a}}$\;
        $O \gets $ assign qubits in $C$ as defined in $\bar{a}_{f}$, repeat reset operations according to $R_p$\;
		\Return $O$
		}
\end{algorithm}
\subsection{Qubit Assignment using Reset Error Characterization} \label{sec mapomatic}

After determining a mapped quantum circuit that can be performed on the target quantum computer in principle, the assignment of the quantum circuit qubits to the physical qubits of the quantum computer is adapted to the current reset error characterization and calibration data provided by the quantum computer operator.

This adaptation is performed with a low classical runtime overhead such that most recent calibration and reset error data can be used.
Furthermore, this step does not incur a further insertion of swap gates: even though a set of physical qubits is determined while computing a feasible assignment to the developed SMT-model (see \cref{sec sat}), current and near-term quantum computers have a regular qubit layout and connectivity.
For instance, the inter-qubit connectivity in \cite{8} is represented by a two-dimensional grid and in \cite{35, 26} by a heavy hex lattice.
This regular qubit layout and connectivity allows a mapped quantum circuit to be placed on different sets of physical qubits on the quantum computer after mapping.

This is exploited by methods such as \cite{28}, where the mapped quantum circuit is assigned to a set of physical qubits that does not incur the insertion of further swap gates and optimizes the cost of a physical qubit assignment as shown in \cref{fig:mapomatic}.
In this work, we extend this approach by considering the reset errors yielded by the reset error characterization experiments and by allowing a repetition of reset operations if this is expected to reduce errors.

Optimizing over the possible qubit assignments of a quantum circuit is performed by the steps in \cref{alg:qubit_assignment}.
First, the calibration data $\xi_{g_{x}}$ is retrieved from the quantum computer operator.
The calibration data $\xi_{g_{x}}$ contains the fidelity of each supported gate $g$ that is applied to the location $x$ in the quantum computer.
The location $x$ can either be a qubit in the quantum computer or the connection between two qubits.
Then, the reset error characterization experiments are executed on the target quantum computer, yielding reset error $\mathcal{R}_{r, p}$ per qubit $p$ and reset repetitions $r$.
In this work, we use the third reset error characterization experiment introduced in \cref{sec char}.
Let $\epsilon_{r, p}$ be the duration of performing $r$ reset operations on qubit $p$, then
\begin{equation}\label{eq res error}
    \xi_{r, p} := e^{-\epsilon_{r, p}/T} \cdot  (1-\mathcal{R}_{r, p})
\end{equation}
indicates the fidelity of repeating the reset operation $r$ times on qubit $p$ including the incurred decoherence on other qubits for a decoherence time $T$ of the target quantum computer.
In a subsequent step, the number of repetitions $R_{p}$ yielding the highest fidelity $\xi_{r, p}$ for qubit $p$ is stored as $R_p$.
Lines 5-7 in \cref{alg:qubit_assignment} are performed by invoking the method introduced in \cite{28} with the cost function:
\begin{equation} \label{eq mapo cost}
    \text{cost}(\bar{a}, \xi, R_{p}) := 1-\prod_{g_{x} \in C} \xi_{g_{x}}\prod_{r_{p}\in C} \xi_{R_{p}, p},
\end{equation}
where $C$ is the mapped quantum circuit $g_x$ is a quantum gate in circuit $C$ applied to location $x$ according to qubit assignment $\bar{a}$ and $r_{p}$ is a reset operation applied to qubit $p$ $R_{p}$ times.
Finally, the qubit assignment $\bar{a}_{f} \in \bar{A}$ with minimal cost $K_{\bar{a}}$ is applied to the mapped target quantum circuit $C$ with reset operations on qubit $p$ repeated $R_p$ times to yield the quantum circuit $O$ with optimized qubit assignment. 

\subsection{Example: 7-Qubit Bernstein-Vazirani with Qubit Reuse}
Given a 7-Qubit Bernstein-Vazirani as the target quantum circuit, a coupling graph of a 27-qubit quantum computer provided by IBM Quantum (see \cref{fig qubit reset reuse}) and the objective to minimize the number of swap gates, the developed method first generates and solves an SMT-model based on these inputs.
The result is a satisfying assignment to the generated SMT-model from which the mapped quantum circuit shown in \cref{fig qubit reset reuse}c) will be derived.

\begin{figure}
    \centering
    \includegraphics[width=\linewidth]{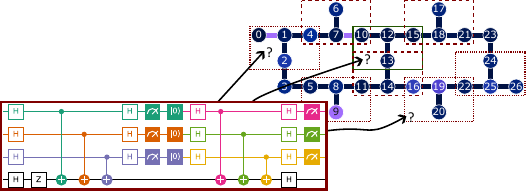}
    
    \caption{Mapping a 7-qubit Bernstein-Vazirani quantum circuit with three reused qubits (left, also see \cref{fig qubit reset reuse}c)) to the coupling graph of $\texttt{ibmq\_ehningen}$ (right).
   }
    \label{fig coupling graph}
    \label{fig:mapomatic}
\end{figure}

\begin{table}
  \caption{Reset error data considered for qubit assignment.}
  \label{tab reset data}
  \includegraphics[width=\linewidth]{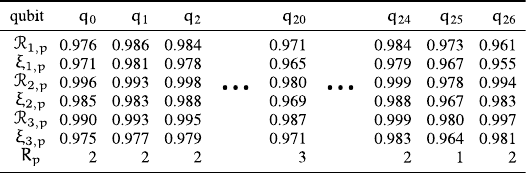}
\end{table}

Then, the steps outlined in \cref{alg:qubit_assignment} are performed.
First, the gate fidelities are extracted from the calibration data by inspection \cite{2} and the third reset error characterization experiment (Reset X (Simul.)) described in \cref{sec char} is performed yielding the data in \cref{tab reset data}.

Given reset operation durations and a decoherence time by the quantum computer operator, applying \cref{eq res error} may yield the values given in rows two, four, and six in \cref{tab reset data}.
From these values, we can extract an optimized number of reset repetitions (last row in \cref{tab reset data}) and an optimized qubit assignment.
The mapped quantum circuit with the minimal number of swap gates in \cref{fig qubit reset reuse} requires four qubits where three qubits are connected to one common qubit.
The coupling graph given in \cref{fig coupling graph} has eight sets of qubits on which this quantum circuit can be assigned without incurring a further insertion of swap gates.
Computing the function given by \cref{eq mapo cost} on each of these qubit sets yields the set of qubits with the least cost, i.e. with the highest expectation of successful computation.
Let the set of qubits $(10, 13, 15, 12)$ yield the lowest cost.
The qubits in the mapped quantum circuit $(0, 1, 2, 3)$ are assigned to $(10, 13, 15, 12)$ respectively, and each reset operation occurring on the assigned qubit is repeated as given by the last row in \cref{tab reset data} before the quantum circuit is submitted to the target quantum computer for computation.

\section{Evaluation of Qubit Reuse Optimization} \label{sec results}
\label{sec method_res}
In this section, the developed qubit reuse optimization method is evaluated on Bernstein-Vazirani (BV) \cite{46} quantum circuits, Hadamard ladder (H-ladder) \cite{17} quantum circuits, and a quantum circuit computing the exclusive-or (XOR) function (xor5\_254) using up to ten qubits \cite{37}.
In contrast to the Bernstein-Vazirani quantum circuit and the exclusive-or quantum circuit, the Hadamard ladder quantum circuit does only require linear qubit connectivity.
Thus, no swap gate insertions are required to compute Hadamard ladder quantum circuits on near-term quantum computers, while the other types of quantum circuits require swap gate insertions depending on the size of the quantum circuit.

We evaluated the developed qubit reuse optimization method on the investigated quantum circuits with two objective functions.
One of the evaluated objective functions minimizes the number of required swap gate insertions while the other objective function minimizes the number of qubits on the quantum computer required by the quantum circuit.
The second objective function is also used to optimize a target quantum circuit using qubit reuse such that its qubit requirement matches a specified number.

The characteristics of each investigated quantum circuit is reported after it was optimized and mapped by the method developed in this work or by qiskit \cite{38}.
The quantum circuits generated by qiskit are used as a basis for comparison.
Among the quantum circuit characteristics evaluated in this work are the quantum circuit depth, number of required swap gate insertions, number of required qubits on the quantum computer, the Hellinger fidelity, and the estimated success probability (ESP).
The estimated success probability (ESP) \cite{5} is computed as the product of fidelities of the quantum gates, measurements, and reset operations occurring in a quantum circuit.
The respective fidelities are obtained by the quantum computer operator during calibration and by the reset error characterization experiments introduced in this work (see \cref{sec char}) that are run shortly before the investigated quantum circuits.
The Hellinger fidelity is determined by the overlap of measurement outcomes determined in error-free simulation and by executing the mapped optimized quantum circuit on a quantum computer.
The 27-qubit IBM quantum computer \texttt{ibm\_hanoi} was used for determining the estimated success probability (ESP) and the Hellinger fidelity of the generated quantum circuits.

\subsection{Impact of Qubit Reuse on Circuit Characteristics}
\cref{tab cdf} shows the impact of quantum circuit mapping and optimization performed by qiskit and the method developed in this work on the quantum circuit characteristics of the investigated quantum circuits.
The data obtained by applying qiskit to the investigated quantum circuits is reported as absolute values in the first four columns of the table.
The remaining columns contain data determined by applying the method developed in this work with various optimization objectives to the investigated quantum circuits.
These remaining columns, except for qubit columns, contain values relative to the data reported by qiskit.
For instance, applying the developed method with the objective to minimize the number of swaps on a Bernstein-Vazirani quantum circuit with 7 qubits (BV7 - second row, columns five to eight) yields a 4-qubit quantum circuit with a 13\% reduced quantum circuit depth, a 20\% improved ESP, and no required swap gate insertions.

\begin{table*}
\centering
\caption{Quantum circuit depth, ESP, as well as the number of qubits and swap gates for the investigated quantum circuits.}
\label{tab cdf}
\begin{tabular}{lrcclrcclrcclrccl}
            & \multicolumn{4}{c}{qiskit (absolute values)} & \multicolumn{4}{c}{mininimal number of swaps} & \multicolumn{4}{c}{one qubit reused} & \multicolumn{4}{c}{most qubit reused} \\
            \cmidrule(lr){2-5}\cmidrule(lr){6-9}\cmidrule(lr){10-13}\cmidrule(lr){14-17}
    circuit & qubits & depth & swap & ESP &   qubits & depth & swap & ESP &    qubits & depth & swap & ESP &       qubits & depth & swap & ESP \\
\midrule
         BV4 &      4 &    10 &    0 &     0.93 &        4 &  1.10x &    0.00x &     1.0x &         3 &  1.70x &    0.00x &     1.00x &            2 &  2.60x &    0.00x &     0.99x \\
         BV7 &      7 &    23 &    4 &     0.70 &        4 &  0.87x & 0.00x &     1.2x &         6 &  0.83x & 0.50x &     1.17x &            2 &  2.30x & 0.00x &     1.20x \\
         BV10 &     10 &    33 &   9 &     0.54 &        4 &  0.85x & 0.00x &    1.4x &         9 &  0.82x & 0.78x &     1.05x &             2 &  2.42x & 0.00x &     1.41x \\
       H-ladder3 &      3 &    8 &    0 &     0.96 &        3 &  1.00x &    0.00x &     1.0x &         2 &  1.88x &    0.00x &     0.99x &            2 &  1.89x &    0.00x &     0.99x \\
       H-ladder5 &      5 &    11 &    0 &     0.93 &        5 &  1.00x &    0.00x &     1.0x &        4 &  1.27x &    0.00x &     0.99x &            2 &  2.18x &    0.00x &     0.95x \\
       H-ladder7 &      7 &    14 &    0 &     0.86 &        7 &  1.00x &    0.00x &     1.0x &        6 &  1.14x &    0.00x &     1.02x &            2 &  2.43x &    0.00x &     0.95x \\
   xor5\_254 &      6 &    10 &    2 &     0.84 &        4 &  1.00x & 0.00x &     1.0x &         5 &  0.80x & 0.50x &     1.03x &            2 &  1.60x & 0.00x &     1.03x \\
\bottomrule
\end{tabular}
\end{table*}

The columns corresponding to the minimization of swap gate insertions in \cref{tab cdf} (columns five to eight) show that the developed method is able to replace the required swap gate insertions of all investigated quantum circuits by suitably selecting qubit reset and reuse at no quantum circuit depth overhead compared to the quantum circuits determined by qiskit.
However, in general, qubit reuse may have no impact on swap gate insertions or may even increase the number of required swap gates insertion.

The largest improvement in ESP can be observed for the 10-qubit Bernstein-Vazirani quantum circuit if the maximum number of qubits are reused (last column, third row in \cref{tab cdf}).
While this also reduces the number of required swap gate insertions, the quantum circuit depth is increased by 142\%, which may lead to larger errors due to decoherence.

Furthermore, reusing one qubit in the 4-qubit Bernstein-Vazirani quantum circuit (BV4) does not reduce the number of required swap gates insertion (row one, columns nine to twelve).
However, the adverse effect of inserting one reset operation in the quantum circuit on the ESP can be completely mitigated by the ability to select a set of qubits for quantum computation that exhibit higher gate fidelities (row 1, column twelve).
The adverse effect of inserting reset operations without replacing swap gate insertions can only be partially mitigated by qubits with higher fidelity, in general.
For instance, reducing the qubit requirement of the 7-qubit or 5-qubit Hadamard ladder quantum circuit to two qubits reduces the ESP by 5\% while more than doubling the quantum circuit depth.

We conclude that while qubit reuse can completely replace the required low-fidelity swap gate insertions and enables to use qubits with higher fidelity for some quantum circuits, it can also have an adverse impact on the characteristics of other quantum circuits.

\subsection{Impact of Increasing Qubit Reuse}
\cref{tab qubit reuse} reports the impact of increasing qubit reuse on the quantum circuit characteristics and Hellinger fidelity of the 10-qubit Bernstein-Vazirani quantum circuit (BV10), 7-qubit Hadamard ladder (H-Ladder7), and the quantum circuit implementing the exclusive-or function (xor5\_254).
For all of the evaluated quantum circuits, there is a pronounced increase in quantum circuit depth with an increase in qubit reuse.
The most significant increase in quantum circuit depth can be observed when decreasing the qubit requirement from three qubits to two qubits where the quantum circuit depth increases by 82\%, 26\%, and 33\% respectively for the BV10, H-Ladder7, and xor5\_254 quantum circuits.
However, by allowing a quantum circuit depth increase of 50\% compared to the quantum circuit generated by qiskit, 7 qubits, 2 qubits, and 3 qubits can be reused using the method developed in this work.

Furthermore, not all possible qubits must be reused for minimizing the number of swap gate insertions. Both the XOR and the Bernstein-Vazirani quantum circuit need no swap gate insertions on the heavy-hex qubit connectivity when the qubit requirement has been lowered to four while the maximum qubit reuse lowers the qubit requirement to two for both quantum circuits.
Note that the 7-qubit Hadamard ladder quantum circuit includes one swap gate insertion when two qubits reused.
This is a consequence of optimizing for quantum circuit depth instead of the number of swap gates after reaching the desired qubit requirement for a quantum circuit.

Using qubit reuse, the developed method improved the Hellinger fidelity by 4.3x for the 10-qubit Bernstein-Vazirani quantum circuit and by 1.16x for the quantum circuit computing the XOR function.
The Hellinger fidelity of the 7-qubit Hadamard ladder quantum circuit could not be improved using qubit reuse.
The impact on the Hellinger fidelity coincidences with the ability of qubit reuse to reduce the number of swap gate insertions in a quantum circuit.
While the change in Hellinger fidelity of the XOR quantum circuit roughly correlates with the change in ESP for increasing qubit reuse, the Hadamard ladder and Bernstein-Vazirani quantum circuits do not exhibit such a pattern.
For the Bernstein-Vazirani quantum circuit, the ESP increases by 40\% compared to the qiskit solution while the Hellinger fidelity improves by 330\%.
For the Hadamard ladder quantum circuit, the ESP decreases by 5\% while the Hellinger fidelity decreases by 93\%.
We suspect this to be caused by two potential effects.
First, the reset error characterization conducted in this work (see \cref{sec char res exp}) reported a large rate of outliers for the reset fidelity.
Thus, the reset operation may fail to realize the desired state transformation occasionally for single quantum circuits even if the average reset fidelity is sufficiently high.
Second, the reset operation may incur additional state manipulations, potentially to neighboring qubits, that are not captured by a metric such as ESP where independent errors are assumed.
In this regard, we refer to \cite{10} where crosstalk errors are reported.

\begin{table}
\addtolength{\tabcolsep}{-0.3em}

\centering
\caption{Hellinger fidelity and circuit characteristics for no reused qubits (qiskit) to 8 reused qubits (this work).}
\begin{tabular}{ clccccccccc }
& & \multicolumn{9}{c}{Reused Qubits} \\
\cmidrule(lr){3-11}
circ. & charact. & 0 & 1 & 2 & 3 & 4 & 5 & 6 & 7 & 8\\
\midrule
\multirow{4}{*}{\rotatebox[origin=c]{90}{BV10}} & depth & 33 & 27 & 24 & 30 & 28 & 30 & 28 & 44 & 80 \\
& swap & 9 & 7 & 5 & 3 & 2 & 1 & 0 & 0 & 0 \\
& ESP & 54\% & 57\% & 49\% & 51\% & 73\% & 75\% & 76\% & 75\% & 76\% \\
& fidelity & 9\% & 18\% & 8\% & 11\% & 5\% & 6\% & 2\% & 12\% & 39\% \\
\midrule
\multirow{4}{*}{\rotatebox[origin=c]{90}{H-Ladder7}} & depth & 14 & 16 & 21 & 27 & 34 & - & - & - & -\\
& swap & 0 & 0 & 1 & 0 & 0 & - & - & - & - \\
& ESP & 86\% & 87\% & 84\% & 83\% & 82\% & - & - & - & -\\
& fidelity & 79\% & 74\% & 9\% & 2\% & 6\% & - & - & - & - \\
\midrule
\multirow{4}{*}{\rotatebox[origin=c]{90}{xor5\_254}} & depth & 10 & 8 & 10 & 12 & 16 & - & - & - & - \\
& swap & 2 & 1 & 0 & 0 & 0 & - & - & - & - \\
& ESP & 84\% & 86\% & 87\% & 88\% & 86\% & - & - & - & -\\
& fidelity & 79\% & 88\% & 92\% & 90\% & 87\% & - & - & - & - \\
\end{tabular}
\label{tab qubit reuse}
\end{table}

\section{Conclusion}\label{sec conclusion}
In this work, we first introduced a set of reset error characterization experiments and quantified the reset fidelity on a 27-qubit quantum computer.
We further introduced a novel quantum circuit optimization method that utilizes qubit reuse to determine a mapped quantum circuit with minimal quantum circuit depth, minimal number of swap gate insertions, or minimal number of qubits required on the quantum computer.
The developed SAT-based approach is augmented by an efficient low-latency qubit assignment that takes the individual reset fidelity of a qubit as well as quantum gate fidelities and measurement fidelities into consideration.
The developed quantum circuit optimization method is then evaluated on a number of quantum circuits with up to 10 qubits to determine optimized quantum circuits ready to be executed on a near-term quantum computer with heavy-hex connectivity.

The introduced reset error characterization experiments revealed a high reset fidelity variance that ranges from 67.5\% to 100\% depending on the reset qubit.
We further identified state-dependent errors during the reset operation and the simultaneous reset of qubits initialized to the $\ket{1}$-state to yield a lower bound on the reset fidelity for quantum circuit optimization purposes.
Using qubit reuse, the developed SAT-based quantum circuit optimization method demonstrated the ability to replace swap gate insertions and an increase in Hellinger fidelity of up to 4.3x.
Furthermore, the evaluation indicated that completely exhausting qubit reuse for quantum circuit optimization may not yield the best quantum circuit characteristics or Hellinger fidelity.
Additional decoherence introduced by an increased quantum circuit depth, as well as the ability to replace swap gate insertions by qubit reuse are an important guide to determining an improved computation performance on a near-term quantum computer.

\section{Acknowledgments}
This work was partially funded by the Carl Zeiss foundation and by the Ministry of Economic Affairs, Labour and Tourism Baden Württemberg in the frame of the Competence Center Quantum Computing Baden-Württemberg (project ‘QORA’).
This material is based upon work supported by the U.S. Department of Energy, Office of Science, National Quantum Information Science Research Centers, Co-design Center for Quantum Advantage (C2QA) under contract number DE-SC0012704.
We acknowledge the use of IBM Quantum services for this work. 
%The views expressed are those of the authors, and do not reflect the official policy or position of IBM or the IBM Quantum team.
\renewcommand*{\bibfont}{\footnotesize}
\renewcommand*{\UrlFont}{\rmfamily}
\printbibliography

\end{document}